\documentstyle[11pt,IAU207_pasp,twoside,psfig]{article}
\def\edcomment#1{\iffalse\marginpar{\raggedright\sl#1\/}\else\relax\fi}
\reversemarginpar

\begin{document}
\title{Age Dating of Globular Clusters Using UBV(RI)\\
Main-sequence Photometry in the Two-color\\
Diagram: Age of NGC 6397}
\author{Valery V. Kravtsov}
\affil{Sternberg Astron. Ins., 13, University Ave., Moscow 119899,
Russia}

\begin{abstract}
I propose and apply a method for deriving ages of the metal-poor globular
clusters (GCs) in a distance-independent way, which is based on age indicator
related to the main sequence in the two--color diagrams with $U-B$ index.
Age of the metal-poor GC NGC 6397 has been estimated, using Yale isochrones
of Demarque et al. (1996), to be close to 16 Gyr provided the cluster
metallicity is near $[Fe/H]=-2.0$.
\end{abstract}

\section{Age Indicator, Isochrones and Observational Data}

The main sequence of metal-poor GCs exhibits a bend in the two--color
diagram(s). Analysis of the multicolor Yale isochrones of Demarque et al.
(1996) shows that color width of such a bend, i.e., a color
difference between the turnoff point and main sequence at the same level of
the $U-B$, is {\bf a function of age at a given metallicity}. Hence age of a
GC can be derived by fitting isochrones of appropriate metallicity to the
cluster two--color diagram(s). The best fit isochrone
must show the most adequate reproduction of fiducial line of the sequence
(mainly in its upper part) and location of its three basic points: points of
the turnoff and main sequence at the same level of the $U-B$, and point of
the extremum.

Multicolor photometry by Alcaino et al. (1997) of NGC 6397 has been used to
apply this approach. The densest part of the cluster field and stars above
the turnoff level were excluded from consideration. Since the most probable
cluster metallicity (see Gratton et al. 2000; Castilho et al. 2000) is close
to $[Fe/H]=-2.0$, the set of isochrones of appropriate metallicity was used.

\section{Result}

Panels of Fig. 1 show that the fitting criteria described above are best met
by isochrone for 16 Gyr age in each diagram presented.
A fitting accuracy is of order $\pm1$ Gyr, taking into account that result
of the best fit is approximately the same in all the diagrams. However,
real error in age may be as large as a few Gyr
(VandenBerg 1999, private communication) even if a photometry is precise,
with no systematic effects. That may be due to incertaincy in
the parameters of theoretical calculations such as $T_{eff}$, or
to the problems of transformation of isochrones to the colors used,
etc. (e.g., see, for details, Grundahl et al. 2000). Note that change of 0.1
dex in $[Fe/H]$ causes change in age estimate $\sim 1$ Gyr so that say
for $[Fe/H]=-1.9$ the age derived for NGC 6397 is $\sim 15$ Gyr.

\begin{figure}
\psfig{figure=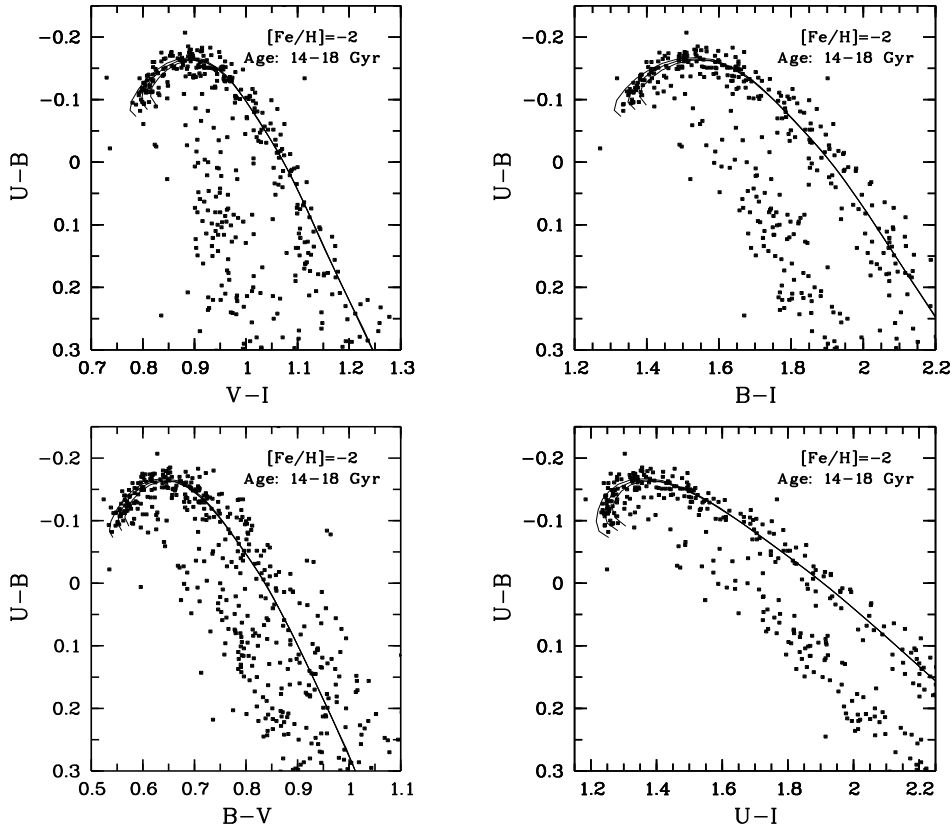,height=11cm,width=12.5cm,angle=270}
\caption{The two--color diagrams of NGC 6397, which show the main sequence
with the superimposed Demarque et al.
(1996) isochrones with an age of 14, 16, and 18 Gyr, $Y=0.23$, $[Fe/H]=-2$.
The panels demonstrate that the fitting criteria are best met in the
diagrams by isochrone for 16 Gyr age. Two other isochrones are shown for
comparison. Stars below the main sequence are field stars.}
\end{figure}

\acknowledgements

I am deeply thankful for the IAU travel grant. Also, I am very grateful to
Dr. D. VandenBerg for valuable discussions and comments.

\end{document}